\documentclass[12pt,a4paper]{article}
    \usepackage{latexsym}
    \usepackage[dvips]{graphicx}
    \usepackage{amssymb,amsmath}
\begin{document}

\title{Geometry of the Hopf Bundle and spin-weighted Harmonics}

    \author{Norbert Straumann\\
        Institute for Theoretical Physics, University of Zurich,\\
        Winterthurerstr. 190, 8057 Zurich, Switzerland} 
    \date{}
    \maketitle
    
\begin{abstract}
We demonstrate that it is conceptually and computationally favorable to regard spin-weighted spherical harmonics as vector valued functions on the total space $SO(3)$ of the Hopf bundle, satisfying a covariance condition with respect to the gauge group $U(1)$ of this bundle. A key role is played by the invariant connection form of the principle Hopf bundle, known to physicists from the geometry behind magnetic monopoles. 
\end{abstract}
    
 \section{Introduction} 
 
Spin-weighted spherical harmonics often play an important role in the mathematical analysis of physical problems, in particular in gravitational physics because of the tensor character of the metric field. A prominent example is the study of the polarization anisotropies of the cosmic microwave background. (For reviews see \cite{RD}, \cite{COM}.)

In contrast to the standard spherical harmonics, spin-weighted spherical ones, with a non-trivial spin weight, can not be considered globally as functions on the 2-sphere. Similar to magnetic monopoles, one has to use at least two patches, together with a $U(1)$-valued transition function, for an appropriate treatment. The Hopf bundle provides again the most suitable geometrical framework for a conceptually satisfactory intrinsic treatment. Moreover, we shall see that some of the rather involved calculations in standard treatments, as for instance in \cite{Gold}, can be avoided.

\section{Some differential geometric tools}

For the benefit of readers not so familiar with the differential geometry of connections in principle fiber bundles, we provide in this section some background material that is used afterwards.

\subsection{Invariant connection in a homogeneous principle fiber bundle}

The Hopf bundle $SO(3)(S^2,SO(2))$ is a special case of a homogeneous principal fiber bundle $G(G/H,H)$ over $M=G/H$ for Lie groups $H\subset G$. We denote their Lie algebras by $\mathcal{H}$ and $\mathcal{G}$, respectively. Let $\theta$ be the Maurer-Cartan form (canonical 1-form) on $G$. We recall that this $\mathcal{G}$-valued 1-form associates by definition to every left-invariant vector field $X$ on $G$ its value $X(e)$ for the unit element $e\in G$. $\theta$ is left-invariant, $L_g^{\ast}\theta=\theta$, and behaves under right-translations $R_g$ according to
\begin{equation}\label{eq2}
R_g^{\ast}\theta=Ad(g^{-1})\theta \quad \mbox{for all} \quad g\in G.
\end{equation}

We assume that the homogeneous space $M=G/H$ is reductive, i.e., that there is a decomposition
\begin{equation}\label{eq1}
\mathcal{G}=\mathcal{H}\oplus \mathcal{M} \quad \mbox{with}\quad Ad(H) \mathcal{M}\subset \mathcal{M}.
\end{equation}
With respect to this direct decomposition, the $\mathcal{H}$-component $A$ of $\theta$ defines a connection on the principal bundle 
$G(G/H,H)$, which is also invariant under left translations by the structure group $H$. This connection of the Hopf bundle will play a key role in our study of spin harmonics.

We give a simple proof of the well-known fact that $A$ is indeed a left-invariant connection form. (This will also fix some of our notation that largely follows that of \cite{KN}).)

a) Let $X\in \mathcal{H}$, and $X^{\sharp}$ the corresponding fundamental vector field on $G,\, X^{\sharp}(g)=dL_g(X)$. Then we have, using the definition of the Maurer-Cartan form $\theta(X^{\sharp})=X$,
\begin{equation*}
A(X^{\sharp})=\theta(X^{\sharp})_{\mathcal{H}}=X.
\end{equation*}

b) From (\ref{eq2})
we deduce the second characterizing property of a connection form
\begin{equation}\label{eq3}
R_h^{\ast} A=Ad(h^{-1})\circ A.
\end{equation}
Indeed, with the decomposition (\ref{eq1}) for which we denote the projection on $\mathcal{H}$ by $pr_{\mathcal{H}}$, we have for a vector field $Y$ on $G$  according to (\ref{eq2})
\[ (R_h^{\ast}A)(Y)=A(R_{h\ast}Y))=  pr_{\mathcal{H}}\circ\theta(R_{h\ast}Y)=pr_{\mathcal{H}}(Ad(h^{-1})\,\theta(Y))=Ad(h^{-1})\, A(Y),\]
where we used in the last step that $Ad(h)$ commutes with the projection $pr_{\mathcal{H}}$, because $\mathcal{H}$ and $\mathcal{M}$ are invariant under $Ad(h)$.
The left-invariance of the connection form $A$ follows from:
\[ L_h^{\ast} A=(L_h^{\ast}\circ  pr_{\mathcal{H}})\theta= (pr_{\mathcal{H}} \circ L_h^{\ast})\theta= pr_{\mathcal{H}}(\theta)=A .\]

\subsection{Properties of the covariant differential associated to $A$}

We first recall the definition of the (exterior) covariant differential $D$ associated to a connection form $A$ on a principal bundle $P(M,H)$. This acts on vector-valued differential $k$-forms $\phi$ on $P$ that are horizontal and satisfy the following covariance condition with respect to the right action $R_h$ (leaving fibers over a point of the base manifold $M$ invariant):
\begin{equation}\label{eq4}
R_h^{\ast} \phi=\rho(h^{-1}) \phi.
\end{equation}
Here $\rho$ is a representation of the structure group $H$ in the target space of $\phi$. By definition, $D\phi$ is the horizontal projection of the exterior differential $d\phi$. This geometrical definition can be translated to the well-known formula
\begin{equation}\label{eq5}
D\phi=d\phi+\rho_{\ast}(A)\wedge \phi,
\end{equation}
where $\rho_{\ast}$ denotes the induced representation of $\mathcal{H}$. A crucial property of $D$ is that it maps horizontal differential $k$-forms of type $\rho$ into horizontal $(k+1)$-forms of the same type $\rho$.

A further property, used later, is that for a left-invariant connection in the homogeneous $H$- principle bundle $G(G/H,H)$ the differential $D$ commutes with $L_g^{\ast}, \,g\in G$:
\begin{equation}\label{eq6}
D\circ L_g^{\ast}=L_g^{\ast} \circ D \quad\mbox{for a left-invariant} \:  A.
\end{equation}
Indeed, from (\ref{eq5}) we obtain
\[ L_g^{\ast}(D\phi)=d(L_g^{\ast}\phi)+L_g^{\ast}(\rho_{\ast}(A))\wedge L_g^{\ast}\phi , \]
and (\ref{eq6}) follows from $L_g^{\ast} A= A$, since $L_g^{\ast}(\rho_{\ast}(A) )= \rho_{\ast}( L_g^{\ast}A)=\rho_{\ast}(A)$. 

At this point we specialize these considerations to the Hopf bundle.

A basis of left-invariant 1-forms on $SO(3)$ in terms of the Euler angles $(\varphi,\vartheta,\psi)$ is given by
\begin{eqnarray}\label{eq14}
\theta^1 &=& -\sin \vartheta\cos \psi\, d\varphi +\sin \psi \,d\vartheta,\nonumber \\
\theta^2 &=& \sin \vartheta\sin \psi \,d\varphi + \cos \psi \,d\vartheta, \nonumber \\
\theta^3 &=& \cos \vartheta\, d\varphi +d\psi.
\end{eqnarray}
The forms ${\theta^i}$ satisfy the Maurer-Cartan equations:
\[d\theta^1 +\theta^2\wedge\theta^3=0, \quad \mbox{and cyclic permutations}.\]
The dual basis of left-invariant vector fields is
\begin{eqnarray}\label{eq15}
e_1 &=& -\frac{\cos \psi}{\sin \vartheta}\,\partial_\varphi +\sin \psi\,\partial_\vartheta +\cot \vartheta\cos \psi\,\partial_\psi, \nonumber \\
e_2 &=& \frac{\sin \psi}{\sin \vartheta}\,\partial_\varphi + \cos \psi\,\partial_\vartheta - \cot \vartheta \sin \psi \,\partial_\psi, \nonumber \\
e_3 &=& \partial_\psi.
\end{eqnarray}
These are the fundamental vector fields on $SO(3)$ generated by right action on the standard basis ${I_k}$ of the Lie algebra of $SO(3)$, defined by the 1-parameter subgroups about the three orthogonal axis of $\mathbb{R}^3$. So $e_k=I_k^{\sharp}$, and the Maurer-Cartan form $\theta$ on $SO(3)$ satisfies
\begin{equation}\label{eq16}
\theta(e_k)=I_k, \quad \mbox{hence} \quad\theta=\sum_{k} \theta^k I_k.
\end{equation}
Obviously, the left-invariant connection form of the Hopf bundle is given by
\begin{equation}\label{eq17}
A=\theta^3=\cos \vartheta\,d\varphi + d\psi.
\end{equation}
The two vector fields $e_1, e_2$ are therefore horizontal. Furthermore, the horizontal linear combinations $(e_1 \mp i e_2)/\sqrt{2}$ are of type $\pm 1$. The same is true for the dual linear combinations $(\theta^1 \pm i \theta^2)/\sqrt{2}$. The vector field $e_3$ is vertical, and is a fundamental vector field of the bundle with $A(e_3)=1$.

For a horizontal form $\phi$ of type (spin) $s$ (i.e., the representation (\ref{eq12}) below) the covariant differential $D$ is given by
\begin{equation}\label{eq18}
D\phi=d\phi+isA\wedge\phi.
\end{equation}
We know that this preserves the type $s$, and that $D$ commutes with $L_g^{\ast},\,g\in SO(3)$, since $A$ is left-invariant.

In passing we note that the curvature $F$ belonging to $A$ is given by $F=dA=-\sin \vartheta\,d\vartheta\wedge d\varphi$, thus projects on the negative of the volume form of $S^2$. This is an example of a monopole field with monopole number (first Chern number) -2.

We now return to the general situation. Consider the matrix elements $\rho_{ik}(g)$ of a representation of $G$ relative to a basis in the representation space $V$, and let
\begin{equation}\label{eq7}
f_{ik}(g)=\rho_{ik}(g^{-1}).
\end{equation}
These functions obviously satisfy the covariance condition (\ref{eq4}) (with the restriction of $\rho$ to $H$), and therefore $D f_{ik}$ is defined. Under a left translation $L_a, \, a\in G$, we find
\[  (L_a^{\ast} f_{ik})(g)=f_{ik}(ag)=\rho_{ik}(g^{-1} a^{-1})=\sum_l \rho_{il}(g^{-1})\rho_{lk}(a^{-1})=\sum_l f_{il}(g) \rho_{lk}(a^{-1}),   \]
thus the transformation law
\begin{equation}\label{eq8}
L^{\ast}_{g^{-1}} (f_{ik})=\sum_{l} f_{il}\,\rho_{lk}(g), \quad g\in G.
\end{equation}
In other words, for fixed $i$ the functions $\lbrace f_{ik}\rbrace$ transform according to the representation $\rho$. From (\ref{eq6}) it follows that the 1-forms $D f_{ik}$ satisfy the \textit{same} transformation law under the left-action by $G$.

Next, we translate the transformation law (\ref{eq8}) to the pull-backs of $f_{ik}$ by a local section $\sigma:\,M\rightarrow G$ of the (non-trivial) principal bundle $ G(M=G/H,H)$. Note that $M$ is naturally a $G$-manifold: $g\in G$ acts on an element $aH\in G/H$ by left multiplication. Consider again a horizontal form $\phi$, satisfying the covariance condition (\ref{eq4}), and let $\phi_\sigma(x):=\phi(\sigma(x))$ (see the commuting diagram below).

\begin{displaymath}
        \begin{picture}(8,3)
    \put(0,2.5){$G$} \put(2.2,0.5){\vector(-2,3){1.1}}
    \put(4,2.5){$\mathcal{H}$} \put(0.8,2.1){\vector(2,-3){1.1}}
    \put(2,0){$G/H$} \put(4,1.2){$\phi_\sigma$}
    \put(3,0.5){\vector(2,3){1.1}}
    \put(1.5,2.7){$\vector(1,0){2}$} \put(1,1){$ \pi $}
    \put(1.9,1.5){$ \sigma$} \put(2.5,2.75){$\phi$}
    \end{picture}
    \end{displaymath}

An element $g\in G$ transforms $\phi_\sigma$ according to $\phi_\sigma(g\cdot x)=\phi\circ\sigma (g\cdot x)$, where $g\cdot x$ denotes the action of $g$ on $x\in M$. We factorize $\sigma(g^{-1}\cdot x)$ as follows
\[  \sigma(g^{-1}\cdot x) =g^{-1} \sigma(x) \Bigl(\underbrace{\sigma (x)^{-1}g\sigma(g^{-1}\cdot x)}_{\in H} \Bigr)=:g^{-1}\sigma(x)\, h_{(g,x)}.\]
Using the covariance property, we obtain
\[\phi_{\sigma}(g^{-1}\cdot x)=\rho( h^{-1}_{(g,x)} )\phi(g^{-1}\sigma(x))=\rho( h^{-1}_{(g,x)} )(L^{\ast}_{g^{-1}}\phi)(\sigma(x)),    \]
so
\begin{equation}\label{eq9}
\phi_{\sigma}(g^{-1}\cdot x)=\rho( h^{-1}_{(g,x)} )(L^{\ast}_{g^{-1}}\phi)_{\sigma}(x).
\end{equation}

We use this equation in the pull-back of (\ref{eq8}), where $\phi$ becomes the column vector $f_{ik}$ for fixed $i$ (the representation $\rho(g)$ affects only the second index). If $f_{ik}$ now denotes for simplicity its pull-back $\sigma^{\ast}f_{ik}=f_{ik}\circ\sigma$, we obtain the same equation, and (\ref{eq9}) becomes
\begin{equation}\label{eq10}
f_{ik}(g^{-1}\cdot x)=\sum_j\rho( h^{-1}_{(g,x)} )_{ij}\sum_l f_{jl}\,\rho_{lk}(g).
\end{equation}
(The representation $\rho$ of $H$ in the first factor on the right affects only the first index of $f_{ik}$.)

\textit{Application to the Hopf bundle}. For the Hopf bundle the irreducible representations of $G=SO(3)$ are  $(2l+1)$-dimensional, usually denoted by $D^{(l)}$. Thus 
\begin{equation}\label{eq11}
f^{(l)}_{sm}=D^{(l)}_{sm}(g^{-1}), \quad g\in SO(3).
\end{equation}
In this case the structure group $H$ is the Abelian group $SO(2)\cong U(1)$, with the 1-dimensional irreducible representations
\begin{equation}\label{eq12}
\rho^{(s)}(h)=e^{is\alpha} h, \quad h=e^{i\alpha},\, s\in \mathbb{Z}.
\end{equation}
Equation (\ref{eq10}) becomes with the notation $h_{(g,x)}=:e^{i\alpha(g,x)}$
\begin{equation}\label{eq13}
f^{(l)}_{sm}(g^{-1}\cdot x)=e^{-is\alpha(g,x)}\sum_{m'} f^{(l)}_{sm'} (x)\,D^{(l)}_{m'm}(g).
\end{equation}
Conversely, it is easy to show that this transformation law uniquely determines $f^{(l)}_{sm}$, up to an $m$-independent normalization constant. This remark will later turn out to be useful.

\section{Applications to spin-weighted vector-valued \\ differential forms}

In this section we apply the previous results to spin-weighted differential forms. Important formulae for spin-harmonics are obtained without much effort.

\subsection{Action of $D$ on spin-weighted functions}

The pull-back of (\ref{eq18}) with the ('standard') local section $\sigma$: polar angles $(\vartheta,\varphi)$ of $S^2 \mapsto (\varphi,\vartheta,\psi=0)$: Euler angles for $SO(3)$, becomes
\[\sigma^{\ast}(D\phi)=d\phi_\sigma+isA_\sigma\wedge\phi_\sigma,  \]
where $\phi_\sigma=\sigma^\ast\phi,\,A_\sigma=\sigma^\ast A=\cos \vartheta\,d\varphi$. For $\sigma^\ast(D\phi)$ we write $D\phi_\sigma$. In what follows in this subsection we drop for simplicity the pull-back index $\sigma$. Then we formally obtain equation (\ref{eq18}), but with the 1-form $A=\cos \vartheta\,d\varphi$ on $S^2$.

The spin-weighted spherical harmonics $_{s}Y_{lm}$ on $S^2$ (minus the poles) are defined by
\begin{equation}\label{eq19}
\Bigl (\frac{4\pi}{2l+1}\Bigr )^{1/2}\,  _{-s}Y_{lm} (n):=f^{(l)}_{sm}\circ\sigma=D^{(l)}_{sm}(\sigma^{-1}(n)).
\end{equation}
We are interested in an explicit form of their covariant differential. This is just a special case of the following formula for a vector-valued function $_{s}\phi$ of type $s$ that follows from (\ref{eq18})
\begin{equation}\label{eq20}
D(_{s}\phi ) = d(_{s}\phi ) +is\cos \vartheta \,_{s}\phi \,d\varphi = \partial_\vartheta\, (_{s}\phi)\,d\vartheta +(\partial_\varphi +is\cos \vartheta)\, _{s}\phi\,d\varphi.
\end{equation}
In terms of the basis $\theta^{\pm}=(d\vartheta \pm i \sin \vartheta\,d\varphi)/\sqrt{2}$, we obtain the important formula
\begin{equation}\label{eq21}
-\sqrt{2} D(_{s}\phi)= {(\partial\!\!\!\!\!\!\;\diagup}^{\ast}\,_{s}\phi) \,\theta^{+} +{(\partial\!\!\!\!\!\!\;\diagup}\,_s\phi)\, \theta^{-},
\end{equation}
with
\begin{eqnarray}\label{eq22}
{\partial\!\!\!\!\!\!\;\diagup} &=& -\Bigl(\partial_\vartheta +\frac{i}{\sin \vartheta}\partial_\varphi \Bigr) +s\cot \vartheta, \nonumber \\
 {\partial\!\!\!\!\!\!\;\diagup}^{\ast} &=& -\Bigl(\partial_\vartheta -\frac{i}{\sin \vartheta}\partial_\varphi \Bigr) -s\cot \vartheta.
\end{eqnarray}

Since $\theta^{(\pm)}$ are of type $\pm 1$ and $D(_{s}Y_{lm} )$ of type $s$, we conclude that ${\partial\!\!\!\!\!\!\;\diagup}\,_{s}Y_{lm}$ is of type $s+1$ and ${\partial\!\!\!\!\!\!\;\diagup}^{\ast}\,_{s}Y_{lm} $ of type $s-1$. (More precisely, one should say that these objects are pull-backs of fields on $SO(3)$ of the stated types.) Together with the remark after equation (\ref{eq13}), it follows that ${\partial\!\!\!\!\!\!\;\diagup}\,_{s}Y_{lm}$ must be proportional to $_{s+1}Y_{lm}$ and ${\partial\!\!\!\!\!\!\;\diagup}^{\ast}\,_{s}Y_{lm} $ proportional to 
$_{s-1}Y_{lm}$, with proportionality constants that are independent of $m$. These can thus be determined by considering the simple special case $m=0$. Using the following two recursion relations of the associated Legendre functions $P^m_l(x)$:
\[  (1-x^2)\frac{d}{dx} P_l^m(x)=(l+1)xP_l^m(x)-(l-m+1) P_{l+1}^m(x),   \]
\[  \sqrt{1-x^2} \frac{dP_l^m}{dx} = P_l^{m+1} - \frac{mx}{\sqrt{1-x^2}} P_l^m \]
(see, e.g., \cite{Ed}), one readily obtains the well-known formulae
\begin{eqnarray}\label{eq23}
{\partial\!\!\!\!\!\!\;\diagup}\,_{s}Y_{lm} &=& \sqrt{(l-s)(l+s+1)}\, _{s+1}Y_{lm}, \nonumber \\
{\partial\!\!\!\!\!\!\;\diagup}^{\ast}\,_{s}Y_{lm} &=& -\sqrt{(l+s)(l-s+1)}\, _{s-1}Y_{lm}.
\end{eqnarray}
We emphasize, that our derivation of these equations does not involve any lengthy calculations.

\subsection{Splicing of the Hopf bundle with the frame bundle}

We denote with $\omega$ the Levi-Civita connection form on the frame bundle $F(S^2,SO(2))$. Relative to the dual $(\pm)$-bases ${\theta^{(\pm)}}$ and ${e^{(\pm)}}$ the matrix of $\omega$ is given by 
\begin{equation}\label{eq24}
 (\omega)=\left(\begin{array}{ll}
                    iA & \quad 0 \\
                    0 & \quad -iA
                    \end{array}\right).
\end{equation}
This can easily be checked: Since the metric is $g=\theta^{(+)}\theta^{(-)}$, i.e., $g_{\pm\pm}=0, \, g_{+-}=g_{-+}=1$, the symmetry condition $\omega_{+-}=-\omega_{-+}$ holds, and the first structure equation is satisfied. 

For a tensor field $T$ on the 2-sphere, with components $T^{\alpha_1\alpha_2...\alpha_p}_{\beta_1\beta_2...\beta_q}=:T^{(\alpha)}_{(\beta)}$, relative to the $(\pm)$-basis, the covariant Levi-Civita derivative $D^{LC}$ has the components
\begin{equation}\label{eq25}
(D^{LC}T)^{(\alpha)}_{(\beta)} = d\,T^{(\alpha)}_{(\beta)} +s\, i\,A \,T^{(\alpha)}_{(\beta)},
\end{equation}
where the integer $s$ is determined as follows: each upper index $\pm$ adds $\pm 1$ to $s$ and each such lower index subtracts $\pm 1$. 
Comparison of (\ref{eq25}) with (\ref{eq18}) and (\ref{eq21}) shows that the directional derivative $\nabla_X$ belonging to the Levi-Civita derivative for a tensor field $T$ is related to that of $D$ as follows:
\begin{equation}\label{eq26}
 (\nabla_{e^{(\pm)}}T)^{(\alpha)}_{(\beta)}= (D_{e^{(\pm)}}T)^{(\alpha)}_{(\beta)}= -\frac{1}{\sqrt{2}}
\left\lbrace
\begin{array}{l}
 {\partial\!\!\!\!\!\!\;\diagup}^{\ast}T^{(\alpha)}_{(\beta)} \\
\llap{}{\partial\!\!\!\!\!\!\;\diagup}T^{(\alpha)}_{(\beta)} \,,
\end{array}\right.
\end{equation}
with $s$ defined in connection with (\ref{eq25}).

Since equation (\ref{eq24}) shows that $D^{LC}$ is closely related to the differential $D$ discussed previously, although the frame bundle over $S^2$ and the Hopf bundle are quite different objects. For this reason it is natural to consider the splicing of the two bundles, with the combined connection $A\oplus\omega$. (A precise definition of this construction is given in \cite{Bl}.)  The corresponding covariant differential will be denoted by $\mathcal{D}$. For a horizontal vector- valued differential form $\phi$ of type $\rho^s\times \rho^{LC}$ of the structure group $U(1)\times U(1)$ we have
\begin{equation}\label{27}
\mathcal{D} = d\phi +(\rho^s\times \rho^{LC})_{\ast} (A\oplus\omega)\wedge\phi =d\phi+[\rho^s_{\ast}(A)\otimes 1 + 1 \otimes\rho^{LC}_{\ast}(\omega)]\wedge\phi.
\end{equation}
The connection (\ref{eq24}) shows for instance that the second term on the right may sometimes vanish. In some special cases only one summand in the square bracket will not be 0. For example, one sees that
\begin{equation*}
\mathcal{D} \,e^{(\pm)}=0,\quad\mathcal{D}\, \theta^{(\pm)}=0.
\end{equation*}
Here is another typical application: Consider a function $f_{\pm s}$ of type $\pm s$. Then
\[   \mathcal{D} (f_{\pm s} \underbrace{e^{(\pm)} \otimes  \cdot\cdot\cdot \otimes e^{(\pm)}}_{\mbox{s factors}} ) =D^{LC} (f_{\pm s}\, e^{(\pm)} \otimes  \cdot\cdot\cdot \otimes e^{(\pm)})=\mathcal{D} (f_{\pm s})\, e^{(\pm)} \otimes \cdot\cdot\cdot \otimes e^{(\pm)} \]
or
\begin{equation}\label{28}
D^{LC} (f_{\pm s}\, e^{(\pm)} \otimes \cdot\cdot\cdot \otimes e^{(\pm)})=(d f_{\pm s} \pm is A f_{\pm s})\,( e^{(\pm)} \otimes \cdot\cdot\cdot \otimes e^{(\pm)}).
\end{equation}
An equivalent relationship was proven in \cite{Gold}. This reference is a classic source for spin weighted spherical harmonics. These are also well described in Appendix 4 of \cite{RD}.


\begin{thebibliography}{10}

\bibitem{RD}
R. Durrer, The cosmic Microwave Background, Cambridge University Press (2008).

\bibitem{COM} 
N. Straumann, From primordial quantum fluctuations to the anisotropies of the cosmic microwave background radiation, 
Ann. Phys. (Leipzig) \textbf{15}, 701-847 (2006). For an updated and expanded version, see: www.vertigocenter.ch/straumann/norbert.

\bibitem{Gold}
J.N. Goldberg et al., J. Math. Phys. \textbf{8}, 2155 (1967).

\bibitem{KN}
 S. Kobayashi and K. Nomizu, Foundations of Differential Geometry, Vol I, Interscience Publishers, John Wiley \& Sons, New York, (1963).
     
\bibitem{Ed}
A.R. Edmonds, Angular Momentum in Quantum Mechanics. Princeton University Press, Princeton (1957); Chapter II.

\bibitem{Bl}
D. Bleecker, Gauge Theory and Variational Principles, Addison-Wesley Publishing Company, INC. (1981).


\end{thebibliography}
\end{document}